\documentclass[preprint,showpacs,showkeys,preprintnumbers,amsmath,amssymb]{revtex4}
\usepackage{graphicx}
\usepackage{dcolumn}
\usepackage{amsmath,amssymb,graphics}
\usepackage{epsfig,subfigure}

\newcommand {\nn}    {\nonumber}

\begin{document}

\title{Fermions on Thick Branes in the Background of Sine-Gordon Kinks}

\author{Yu-Xiao Liu$^1$} \email{liuyx@lzu.edu.cn}
 \thanks{Corresponding author}
\author{Li-Da Zhang$^1$} \email{zhangld04@lzu.cn}
\author{Li-Jie Zhang$^2$}\email{lijzhang@shu.edu.cn}
\author{Yi-Shi Duan$^1$} \email{ysduan@lzu.edu.cn}

\affiliation{$^1$Institute of Theoretical Physics,
  Lanzhou University, Lanzhou 730000, P. R. China}
\affiliation{$^2$Department of Physics, Shanghai University,
     Shanghai 200444, P. R. China}

\begin{abstract}
A class of thick branes in the background of sine-Gordon kinks
with a scalar potential $V(\phi)=p(1+\cos\frac{2\phi}{q})$ was
constructed by R. Koley and S. Kar [Classical Quantum Gravity
\textbf{22}, 753 (2005)]. In this paper, in the background of the
warped geometry, we investigate the issue of localization of spin
half fermions on these branes in the presence of two types of
scalar-fermion couplings: $\eta\bar{\Psi}\phi\Psi$ and
$\eta\bar{\Psi}\sin\phi \Psi$. By presenting the mass-independent
potentials in the corresponding Schr\"{o}dinger equations, we
obtain the lowest Kaluza--Klein (KK) modes and a continuous
gapless spectrum of KK states with $m^2>0$ for both types of
couplings. For the Yukawa coupling $\eta\bar{\Psi}\phi\Psi$, the
effective potential of the right chiral fermions for positive $q$
and $\eta$ is always positive, hence only the effective potential
of the left chiral fermions could trap the corresponding zero
mode. This is a well-known conclusion which had been discussed
extensively in the literature. However, for the coupling
$\eta\bar{\Psi}\sin\phi \Psi$, the effective potential of the
right chiral fermions for positive $q$ and $\eta$ is no longer
always positive. Although the value of the potential at the
location of the brane is still positive, it has a series of wells
and barriers on each side, which ensures that the right chiral
fermion zero mode could be trapped. Thus we may draw the
remarkable conclusion: for positive $\eta$ and $q$, the potentials
of both the left and right chiral fermions could trap the
corresponding zero modes under certain restrictions.
\end{abstract}

\pacs{11.10.Kk., 04.50.-h. }

\keywords{Large Extra Dimensions,
        Field Theories in Higher Dimensions}


\maketitle


\section{Introduction}

The suggestion that extra dimensions may not be compact
\cite{rs,Lykken,RubakovPLB1983136,RubakovPLB1983139,Akama1983,VisserPLB1985,Randjbar-DaemiPLB1986,StojkovicCHM}
or large \cite{AntoniadisPLB1990,ADD} can provide new insights for
solving gauge hierarchy problem \cite{ADD} and cosmological
constant problem
\cite{RubakovPLB1983139,Randjbar-DaemiPLB1986,KehagiasPLB2004},
etc. In the framework of brane scenarios, gravity is free to
propagate in all dimensions, while all the matter fields are
confined to a 3--brane
\cite{RubakovPLB1983139,VisserPLB1985,ADD,SquiresPLB1986,gog}. In
Ref. \cite{rs}, an alternative scenario of the compactification
had been proposed. In this scenario, the internal manifold does
not need to be compactified to the Planck scale any more, which is
one of reasons why this new compactification scenario has
attracted so much attention. Among all of the brane world models,
there is an interesting and important model in which extra
dimensions comprise a compact hyperbolic manifold
\cite{StojkovicCHM}. The model is known to be free of usual
problems that plague the original Arkani-Hamed-Dimopoulos-Dvali
(ADD) models and share many common features with Randall-Sundrum
(RS) models.

In the brane world scenario, an important question is localization
of various bulk fields on a brane by a natural mechanism. It is
well known that massless scalar fields \cite{BajcPLB2000} and
graviton \cite{rs} can be localized on branes of different types,
and that spin 1 Abelian vector fields can not be localized on the
RS brane in five dimensions, but can be localized in some
higher-dimensional cases \cite{OdaPLB2000113}. Spin 1/2 fermions
do not have normalizable zero modes and hence can not be localized
in five and six dimensions
\cite{BajcPLB2000,OdaPLB2000113,NonLocalizedFermion,IchinosePRD2002,Ringeval,GherghettaPRL2000,Neupane}.

Recently, an increasing interest has been focused on the study of
thick brane scenarios based on gravity coupled to scalars in
higher dimensional space-time
\cite{dewolfe,gremm,Csaki,CamposPRL2002,varios,ThickBraneDzhunushaliev,ThickBraneBazeia}.
A virtue of these models is that the branes can be obtained
naturally rather than introduced by hand \cite{dewolfe}. Besides,
these scalar fields provide the ``material" of which the thick
branes are made. In Ref. \cite{KoleyCQG2005}, exact solutions of
the Einstein--scalar equations with a sine-Gordon potential and a
negative cosmological constant were constructed. In this system
the scalar field configuration in fact is a kink, which provides a
thick brane realization of the brane world as a domain wall in the
bulk. The warped background space-time has a non--constant but
asymptotically negative Ricci curvature. Such a configuration  was
also illustrated in several examples in the literature
{\cite{IchinosePRD2002,Ringeval}}.

The localization problem of spin half fermions on thick branes is
interesting and important. Localization of fermions in general
space-times had been studied for example in
\cite{RandjbarPLB2000}. In five dimensions, with the
scalar--fermion coupling, there may exist a single bound state and
a continuous gapless spectrum of massive fermion KK states
\cite{ThickBrane1,ThickBrane2,ThickBrane3,Liu0708}, while for some
other brane models, there exist finite discrete KK states (mass
gap) and a continuous gapless spectrum starting at a positive
$m^2$ \cite{ThickBrane4,Liu0803}. In Ref. \cite{DubovskyPRD2000},
it was found that fermions can escape into the bulk by tunnelling,
and the rate depends on the parameters of the scalar potential. In
Ref. \cite{KoleyCQG2005}, the authors obtained trapped discrete
massive fermion states on the brane, which in fact are quasi-bound
and have a finite probability of escaping into the bulk. It is
also interesting to note that in Ref. \cite{0803.1458} fermion
modes in a sine-Gordon kink and kink-anti-kink system were also
studied in some 1+1 dimensional scalar field theories. It was
shown that there exist discrete bound states. However, when the
wall and anti-wall approach each other, the system cannot support
fermion bound states and the discrete states merge into the
continuous spectrum of the Dirac equation.

It is known that, under the Yukawa coupling
$\eta\bar{\Psi}\phi\Psi$, only one of the effective potentials of
the left and right chiral fermions could trap the corresponding
zero mode for positive $q$ and $\eta$. In this paper, we will
reinvestigate the localization issues of fermions on the branes
obtained in Ref. \cite{KoleyCQG2005} in the presence of different
types of scalar-fermion couplings, by presenting the
mass-independent potentials in the corresponding Schr\"{o}dinger
equations. We will show that, for the coupling
$\eta\bar{\Psi}\sin\phi \Psi$, not only the potential of the left
chiral fermions but also the potential of the right ones could
trap the corresponding zero modes for positive $\eta$ and $q$
under certain different restrictions for each case. Besides,
instead of discrete massive KK mode, there exists a continuous
gapless spectrum of KK states with $m^2>0$. The shapes of the
potentials also suggest that the massive KK modes asymptotically
turn into plane waves, which represent delocalized massive KK
fermions.

The paper is organized as follows: In section \ref{SecModel}, we
first give a brief review of the thick brane arising from a
sine-Gordon potential in a 5-dimensional space-time. Then, in
section \ref{SecLocalize}, we study localization of spin half
fermions on the thick brane with two different types of
scalar-fermion interactions by presenting the shapes of the
potentials of the corresponding Schr\"{o}dinger problem. Finally,
a brief conclusion and discussion are presented.

\section{Review of the sine-Gordon kink and the thick branes}
\label{SecModel}

Let us consider thick branes arising from a real scalar field with
a sine-Gordon potential
\begin{equation}
 V(\phi)=p\left (1+ \cos \frac{2\phi}{q}\right ).
 \label{Vphi}
\end{equation}
This special potential was considered in Ref. \cite{KoleyCQG2005}
and other different choices of $V(\phi)$ can be found in the work
of others {\cite{gremm}}. In the model, the bulk sine-Gordon
potential provides a thick brane realization of the
Randall--Sundrum scenario, and the soliton configuration of the
scalar field dynamically generate the domain wall configuration
with warped geometry. The action for such a system is given by
\begin{equation}
S = \int d^5 x \sqrt{-g}\left [ \frac{1}{2\kappa_5^2}\left
(R-2\Lambda\right ) -\frac{1}{2} g^{MN}\partial_M \phi
\partial_N \phi - V(\phi) \right ],
\label{action}
\end{equation}
where $\kappa_5^2=8 \pi G_5$ with $G_5$ the 5-dimensional Newton
constant, and $\Lambda$ is the 5-dimensional cosmological
constant. The line-element which results in a 4-dimensional
Poincar$\acute{e}$ invariance of the action (\ref{action}) is
assumed as
\begin{equation}
\label{linee} ds^2=\text{e}^{2A(y)}\eta_{\mu\nu}dx^\mu dx^\nu +
dy^2,
\end{equation}
where $\text{e}^{2A(y)}$ is the warp factor and $y$ stands for the
extra coordinate. The scalar field is considered to be a function
of $y$ only. The field equations, that are derivable from
(\ref{action}) with the ansatz (\ref{linee}), reduce to the
following coupled nonlinear differential equations
\begin{eqnarray}
A '' & = & -\frac{\kappa_5^2}{3} {\phi'}^2, \\
{A'}^2 & = & -\frac{\kappa_5^2}{12} \left ({\phi'}^2 - 2V \right )
            +\frac{\Lambda}{6},\\
\phi'' +4A'\phi' &  = & \frac{dV}{d\phi}.
\end{eqnarray}
For the sine-Gordon potential (\ref{Vphi}), the solution can be
calculated \cite{KoleyCQG2005}:
\begin{eqnarray}
 A(y) &=&- \tau \ln
\cosh ky, \label{Ay} \\
 \phi(y) &=& 2q
\arctan\left ( \exp ky \right ) -\frac{\pi q}{2},
 \label{SGKink}
\end{eqnarray}
where $\tau$ and $k$ are given by
\begin{equation}
 \tau=\frac{1}{3}\kappa_5^2 q^2, ~~~
 k=\frac{\sqrt{6|\Lambda|}}{6\tau}. \label{tau_k}
\end{equation}
The parameters $q$ and $\Lambda$ are free to choose, and $p$ is
given by
\begin{equation}
 p= \frac{\vert \Lambda\vert}{2\kappa_5^4} \left
(\frac{\kappa_5^2}{3}+\frac{1}{4q^2}\right ).
\end{equation}
Observing the forms of (\ref{Ay}) and (\ref{SGKink}), one can find
that the configuration of $\phi(y)$ is a kink for positive $q$,
and the warp factor $A(y)$ is a smooth function. Besides these
properties, more detailed discussions can be found in Ref.
\cite{KoleyCQG2005}.

The extensive work had been done on non-supersymmetric
{\cite{dewolfe,Csaki,CamposPRL2002,IchinoseCQG2001}} as well as
supersymmetric {\cite{susy,Maru2001,Maru2003}} domain walls in
different models. In Ref. {\cite{Maru2001}}, Maru {\em et al}
constructed an analytic non-Bogomol'nyi-Prasad-Sommerfeld (BPS)
solution of the sine-Gordon domain wall in a 4-dimensional global
supersymmetric model. In Ref. {\cite{Maru2003}}, this sine-Gordon
domain wall solution was extended to a solution in 4-dimensional
supergravity and its stability had been examined. In these papers,
although sine-Gordon domain wall had been considered in
supersymmetric theories, the properties of the solution itself are
also valid in the purely bosonic sector. In the following, we will
reconsider the issue of localization of spin half fermions on the
3--brane in the presence of two types of kink-fermion couplings in
the background of the sine-Gordon kink (\ref{SGKink}) and the
corresponding warped geometry.

\section{Localization of fermions on the thick branes}
\label{SecLocalize}

Now, let us investigate whether spin half fermions can be
localized on the brane. We will analyze the spectrum of fermions
for the thick branes by presenting the mass-independent potentials
in the corresponding Schr\"{o}dinger equations. In order to get
the mass-independent potentials, we will follow Ref. \cite{rs} and
change the metric given in (\ref{linee}) to a conformally flat one
\begin{equation}
\label{conflinee2} ds_5^2=\text{e}^{2A}\left(\eta_{\mu\nu}dx^\mu
dx^\nu+dz^2\right)
\end{equation}
by performing the coordinate transformation
\begin{equation}
dz=\text{e}^{-A(y)}dy. \label{transformation}
\end{equation}

In five dimensions, fermions are four component spinors and their
Dirac structure is described by $\Gamma^M= e^M _{\bar{M}}
\Gamma^{\bar{M}}$ with $\{\Gamma^M,\Gamma^N\}=2g^{MN}$. In this
paper, $\bar{M}, \bar{N}, \cdots$ denote the local Lorentz indices,
and $\Gamma^{\bar{M}}$ are the flat gamma matrices in five
dimensions. In our set-up,
$\Gamma^M=(\text{e}^{-A}\gamma^{\mu},\text{e}^{-A}\gamma^5)$, where
$\gamma^{\mu}$ and $\gamma^5$ are the usual flat gamma matrices in
the Dirac representation. The Dirac action of a massless spin 1/2
fermion coupled to the scalar is
\begin{eqnarray}
S_{1/2} = \int d^5 x \sqrt{-g} \left(\bar{\Psi} \Gamma^M D_M
\Psi-\eta \bar{\Psi} F(\phi) \Psi\right), \label{DiracAction}
\end{eqnarray}
where the covariant derivative $D_M$ is defined as $D_M\Psi =
(\partial_M + \frac{1}{4} \omega_M^{\bar{M} \bar{N}}
\Gamma_{\bar{M}} \Gamma_{\bar{N}} ) \Psi$ with the spin connection
$\omega_M= \frac{1}{4} \omega_M^{\bar{M} \bar{N}} \Gamma_{\bar{M}}
\Gamma_{\bar{N}}$ and
\begin{eqnarray}
 \omega_M ^{\bar{M} \bar{N}}
   &=& \frac{1}{2} {e}^{N \bar{M}}(\partial_M e_N ^{\bar{N}}
                      - \partial_N e_M ^{\bar{N}}) \nn \\
   &-& \frac{1}{2} {e}^{N \bar{N}}(\partial_M e_N ^{\bar{M}}
                      - \partial_N e_M ^{\bar{M}})  \nn \\
   &-& \frac{1}{2} {e}^{P \bar{M}} {e}^{Q \bar{N}} (\partial_P e_{Q
{\bar{R}}} - \partial_Q e_{P {\bar{R}}}) {e}^{\bar{R}} _M.
\end{eqnarray}
With the metric (\ref{conflinee2}), the non-vanishing components
of the spin connection $\omega_M$ are
\begin{eqnarray}
  \omega_\mu =\frac{1}{2}(\partial_{z}A) \gamma_\mu \gamma_5. \label{eq4}
\end{eqnarray}
Then the equation of motion is given by
\begin{eqnarray}
 \left\{ \gamma^{\mu}\partial_{\mu}
         + \gamma^5 \left(\partial_z  +2 \partial_{z} A \right)
         -\eta\; \text{e}^A F(\phi)
 \right \} \Psi =0, \label{DiracEq1}
\end{eqnarray}
where $\gamma^{\mu} \partial_{\mu}$ is the Dirac operator on the
brane.

Now we study the above 5-dimensional Dirac equation. From the
equation of motion (\ref{DiracEq1}), we will search for the
solutions of the general chiral decomposition
\begin{equation}
 \Psi(x,z) = \sum_n\psi_{Ln}(x) \alpha_{Ln}(z)
 +\sum_n\psi_{Rn}(x) \alpha_{Rn}(z)
\end{equation}
with $\psi_{Ln}(x)=-\gamma^5 \psi_{Ln}(x)$ and
$\psi_{Rn}(x)=\gamma^5 \psi_{Rn}(x)$ the left-handed and
right-handed components of a 4-dimensional Dirac field, the sum over
$n$ can be both discrete and continuous. Here, we assume that
$\psi_{L}(x)$ and $\psi_{R}(x)$ satisfy the 4-dimensional massive
Dirac equations
$\gamma^{\mu}\partial_{\mu}\psi_{Ln}(x)=m_n\psi_{R_n}(x)$ and
$\gamma^{\mu}\partial_{\mu}\psi_{Rn}(x)=m_n\psi_{L_n}(x)$. Then
$\alpha_{L}(z)$ and $\alpha_{R}(z)$ satisfy the following coupled
equations
\begin{subequations}
\begin{eqnarray}
 \left \{ \partial_z+2\partial_{z}A
                  + \eta\;\text{e}^A F(\phi) \right \} \alpha_{Ln}(z)
  &=&  ~~m_n \alpha_{Rn}(z), \label{CoupleEq1a}  \\
 \left \{ \partial_z+2\partial_{z}A
                  - \eta\;\text{e}^A F(\phi) \right\} \alpha_{Rn}(z)
  &=&  -m_n \alpha_{Ln}(z). \label{CoupleEq1b}
\end{eqnarray}\label{CoupleEq1}
\end{subequations}
In order to obtain the standard 4-dimensional action for the
massive chiral fermions, we need the following orthonormality
conditions for $\alpha_{L_{n}}$ and $\alpha_{R_{n}}$:
\begin{eqnarray}
 \int_{-\infty}^{\infty} \text{e}^{4A}  \alpha_{Lm} \alpha_{Rn}dz
   &=& \delta_{LR}\delta_{mn}. \label{orthonormality}
\end{eqnarray}

Defining $\widetilde{\alpha}_{L}=\text{e}^{2A}\alpha_{L}$, we get
the Schr\"{o}dinger-like equation for the left chiral fermions
\begin{eqnarray}
  [-\partial^2_z + V_L(z) ]\widetilde{\alpha}_{Ln}
  =m_n^2 \widetilde{\alpha}_{Ln},
  \label{SchEqLeftFermion}
\end{eqnarray}
where the effective potential is given by
\begin{eqnarray}
  V_L(z)= \text{e}^{2A} \eta^2 F^2(\phi)
     - \text{e}^{A} \eta\; \partial_z F(\phi)
     - (\partial_{z}A) \text{e}^{A} \eta F(\phi).
\end{eqnarray}
For the right chiral fermions, the corresponding potential can be
written out easily by replacing $\eta\rightarrow -\eta$ from above
potential
\begin{eqnarray}
  V_R(z)= \text{e}^{2A} \eta^2 F^2(\phi)
     + \text{e}^{A} \eta\; \partial_z F(\phi)
     + (\partial_{z}A) \text{e}^{A} \eta F(\phi).
\end{eqnarray}
It can be seen clearly that, for the left (right) chiral fermion
localization, there must be some kind of Yukawa coupling. This
situation can be compared with the one in the RS framework
\cite{BajcPLB2000}, where additional localization method
\cite{JackiwPRD1976} was introduced for spin 1/2 fields.
Furthermore, $F(\phi(z))$ must be an odd function of $\phi(z)$
when we demand that $V_L(z)$ or $V_R(z)$ is $Z_2$-even with
respect to the extra dimension $z$. In this paper, we consider two
cases $F(\phi(z))=\phi(z)$ and $F(\phi(z))=\sin\phi(z)$ as
examples. For each case, we get a continuous spectrum of KK modes
with positive $m^2>0$. However, it is shown that only the massless
chiral modes could be localized on the brane.

\subsection{$F(\phi)=\phi$}

\begin{figure}[htb]
\begin{center}
\includegraphics[width=7.5cm,height=5.5cm]{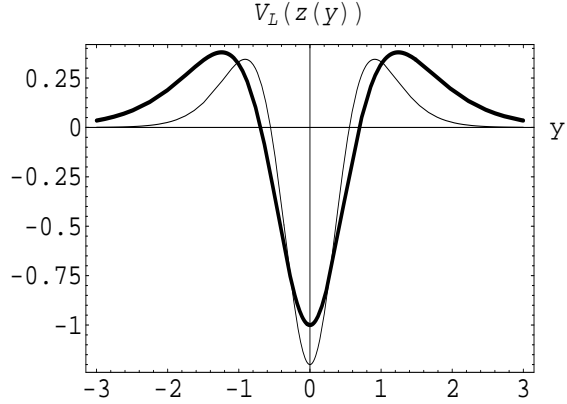}
\end{center}
\caption{The shape of the potential $V_L(z(y))$ for the case
$F(\phi)=\phi$. The parameters are set to $k=1$, $q=1$, $\eta=1$
and $\tau=1$ for thick line, and $\eta=1.2$ and $\tau=2$ for thin
line.}
 \label{fig_Vy1}
\end{figure}

Here, we face the difficulty that for general $\tau$ we can not
obtain the function $y(z)$ in an explicit form. But we can write
the potentials as a function of $y$:
\begin{eqnarray}
  V_L(z(y))
     &=& \frac{1}{4} q \eta  \cosh^{-2\tau} (ky)
      \bigg(-\frac{8\;\text{e}^{ky}k}{1+\text{e}^{2ky}}\nonumber  \\
     &&     +q\eta (\pi-4\arctan \text{e}^{ky})^2   \nonumber  \\
     &&     -2k\tau(\pi-4\arctan \text{e}^{ky}) \tanh ky\bigg), \\
  V_R(z(y))&=& V_L(z(y))|_{\eta\rightarrow -\eta}.
\end{eqnarray}
This potential for the left chiral fermions has the asymptotic
behavior: $V_L(y=\pm \infty)=0$ and $V_L(y=0)=-k q \eta$, where
$k>0$. For $q\eta>0$, this is in fact a volcano type potential
\cite{volcano,Davoudiasl}. The effective potential for the left
chiral fermions is shown in Fig. \ref{fig_Vy1}. For positive
$\tau$, $z(y)$ is a monotonous function, which means that the
potential for arbitrary positive $\tau$ provides no mass gap to
separate the zero mode from the excited KK modes.

\begin{figure}[htb]
\includegraphics[width=7.5cm,height=5.5cm]{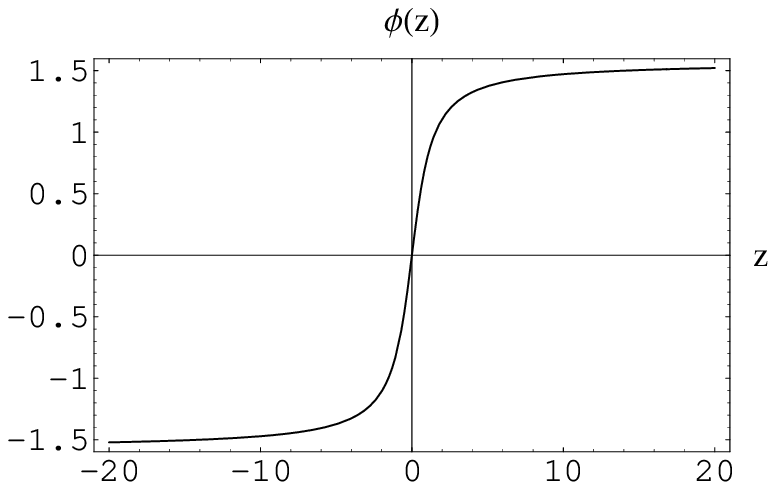}
\includegraphics[width=7.5cm,height=5.5cm]{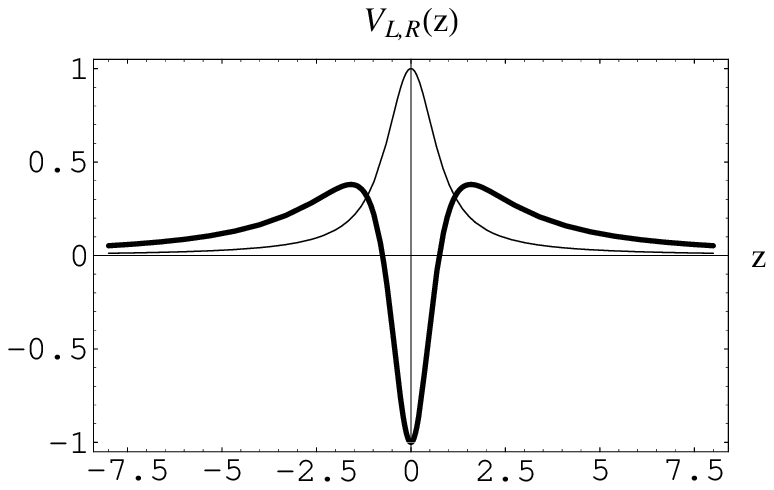}
\caption{The shape of the kink ($\phi(z)$ with positive $k$) and
the potentials $V_L(z)$ (thick line) and $V_R(z)$ (thin line) for
the left and right chiral fermions for the case $F(\phi)=\phi$ and
$\tau=1$ in $z$ coordinate. The parameters are set to $k=1,q=1$,
and $\eta=1$.}
 \label{fig_V_fermion}
\end{figure}

In the following, without loss of generality, we mainly discuss
the case $\tau=1$, for which one can invert the coordinate
transformation $dz=\text{e}^{-A(y)}dy$, namely
\begin{eqnarray}
y=\text{arcsinh}(kz)/k,
\end{eqnarray}
and get the explicit forms of the potentials and the kink
$\phi(z)$
\begin{eqnarray}
 V_L(z)
  &=&\eta \bigg(
   \frac{q^2 \eta (\pi-4\arctan \text{e}^{\text{arcsinh} kz})^2}
         {4(1+k^2 z^2)} \nonumber \\
  &&-\frac{2kq \;\text{e}^{\text{arcsinh} kz} }
         {(1+k^2 z^2)(1+\text{e}^{2\;\text{arcsinh} kz}) }
         \nonumber \\
  &&-\frac{k^2 q z (\pi-4\arctan \text{e}^{\text{arcsinh} kz})}
         {2(1+k^2 z^2)^{3/2}}
  \bigg),  \label{VeffLeftFermion_phi} \\
  V_R(z) &=& V_L(z)|_{\eta \rightarrow -\eta},  \\
  \phi(z) &=& 2q \arctan \text{e}^{\text{arcsinh}(kz)}
             -\frac{\pi q}{2}. \label{phi_z}
\end{eqnarray}
The values of the potentials for the left chiral and right chiral
fermions at $y = 0$ are given by
\begin{equation}
V_R(0) =-V_L(0) =  k q \eta.
\end{equation}
Both potentials have the asymptotic behavior: $V_{L,R}(z=\pm
\infty)=0$. But for a given coupling constant $\eta$, the values
of the potentials at $z=0$ are opposite. The shape of the kink
$\phi(z)$ and the above two potentials are shown in Fig.
\ref{fig_V_fermion} for given values of positive $\eta$ and $q$.
It can be seen that $V_L(z)$ is indeed a volcano type potential.
Hence, the potential provides no mass gap to separate the fermion
zero mode from the excited KK modes, and there exists a continuous
gapless spectrum of the KK modes for both the left chiral and
right chiral fermions.

For positive $q$ and $\eta$, only the potential for left chiral
fermions has a negative value at the location of the brane, which
could trap the left chiral fermion zero mode solved from
(\ref{CoupleEq1a}) by setting $m_0=0$:
\begin{equation}
 \widetilde{\alpha}_{L0}(z) =\text{e}^{2A} {\alpha}_{L0}(z)
 \propto \exp\left(-\eta\int^z dz'\text{e}^{A(z')}\phi(z')\right).
  \label{zeroMode1}
\end{equation}
In order to check the normalization condition
(\ref{orthonormality}) for the zero mode (\ref{zeroMode1}), we
need to check whether the inequality
\begin{equation}
\int dz \exp\left(-2\eta\int^z dz'
  \text{e}^{A(z')}\phi(z')\right)
  < \infty     \label{condition1}
\end{equation}
is satisfied. For the integral $\int dz\text{e}^{A}\phi$, we only
need to consider the asymptotic characteristic of the function
$\text{e}^{A}\phi$ for $z \rightarrow \infty$. Noting that
$\arctan z \rightarrow \pi/2$ when $z \rightarrow \infty$, we have
\begin{eqnarray}
 &&\text{e}^{A}\phi
 =\frac{4q\arctan \text{e}^{\text{arcsinh}kz}-q\pi}
       {2\sqrt{1+k^2 z^2}}
  \rightarrow
  \frac{q\pi}{2\sqrt{1+k^2 z^2}},~~~ \\
 &&\int dz\text{e}^{A}\phi \rightarrow
    \frac{q\pi}{2k} \text{arcsinh}kz.
\end{eqnarray}
Now, the normalization condition (\ref{condition1}) is changed to
$\int dz \exp\left(- \frac{\eta q\pi}{k}\text{arcsinh}kz\right) <
\infty$. Hence the condition on the free parameters $\eta$ and $q$
is
\begin{equation}
\eta q> \frac{k}{\pi}.   \label{condition2}
\end{equation}
In fact, we can solve the problem in $y$ coordinate easily. In this
coordinate, the condition (\ref{condition1}) becomes
\begin{equation}
\int dy \exp\left(-A(y)-2\eta\int^y dy'\phi(y')\right)
  < \infty.     \label{condition3}
\end{equation}
When $y \rightarrow \infty$, we have $A(y)\rightarrow -ky$ and
$\phi(y)\rightarrow q\pi/2$, and so $\left(-A(y)-2\eta\int^y
dy'\phi(y')\right) \rightarrow (k-\eta q\pi)y$. Then we can get
the restriction condition (\ref{condition2}) for localizing the
zero mode of the left chiral fermions.

\begin{figure}[h]
 \centering
 \subfigure[Zero mode ($m^2=0$)]{\label{fig_KKmodes_phia}
  \includegraphics[width=7cm,height=5cm]{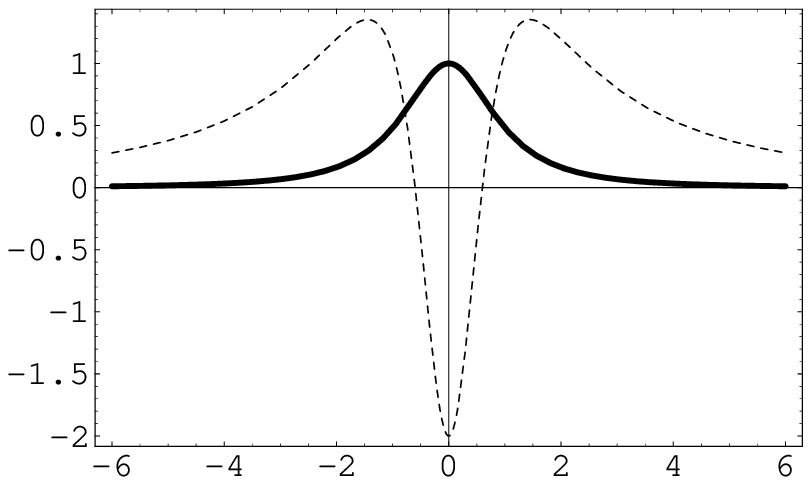}}
 \subfigure[Massive modes ($m^2=1,10$)]
{\label{fig_KKmodes_phib}
  \includegraphics[width=7cm,height=5cm]{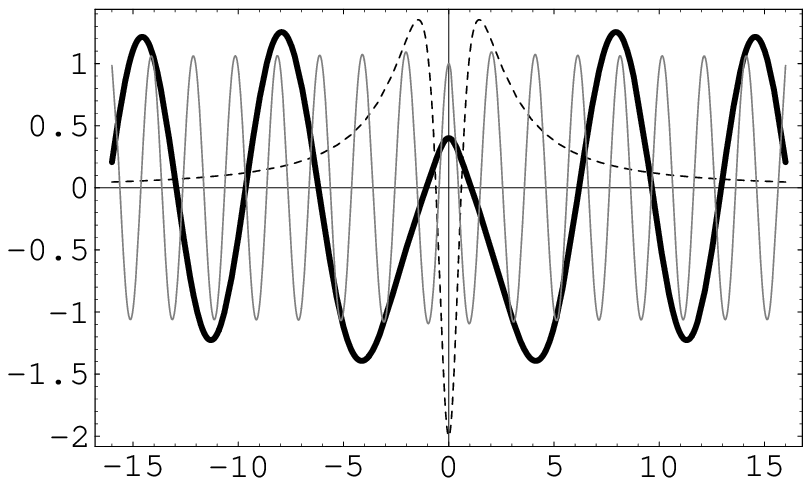}}
 \caption{The shape of the potentials $V_L(z)$
   (\ref{VeffLeftFermion_phi}) (dashed lines),
   the zero mode (\ref{zeroMode1}) and the massive modes
   for the left chiral fermions for the case $F(\phi)=\phi$.
   The parameters are set to $k=1,q=2$, and $\eta=1$.
   In the right figure, we set $m^2=1$ and 10 for thick black line
   and thin gray line, respectively.} \label{fig_KKmodes_phi}
\end{figure}

The zero mode (\ref{zeroMode1}) represents the lowest energy
eigenfunction (ground state) of the Schr\"{o}dinger equation
(\ref{SchEqLeftFermion}) since it has no zeros, and it is the only
one bound state. Since the ground state has the lowest mass square
$m_0^2=0$, there is no tachyonic left chiral fermion mode. The
potential (\ref{VeffLeftFermion_phi}) provides no mass gap to
separate the fermion zero mode from the excited KK modes. In Fig.
\ref{fig_KKmodes_phi}, we plot the left chiral fermion potential
$V_L(z)$, the corresponding zero mode, and the massive KK modes.
We see that the zero mode is bound on the brane, while the massive
modes propagate along the extra dimension. Those massive modes
with lower energy experience an attenuation due to the presence of
the potential barriers near the location of the brane.

In the case $q\eta>0$, the potential for the right chiral fermions
is always positive, which shows that it can not trap the right
chiral zero mode. But for the case of negative $q\eta$, things are
opposite and only the right chiral zero mode can be trapped on the
brane. For arbitrary $q\eta\neq 0$, the two potentials suggest
that there is no mass gap but a continuous spectrum of KK modes
with $m^2>0$.

\subsection{$F(\phi)=\sin\phi$}

\begin{figure}[h]
 \centering
 \subfigure[$q=9.5$]{\label{fig_Fsinphia}
  \includegraphics[width=7cm,height=5cm]{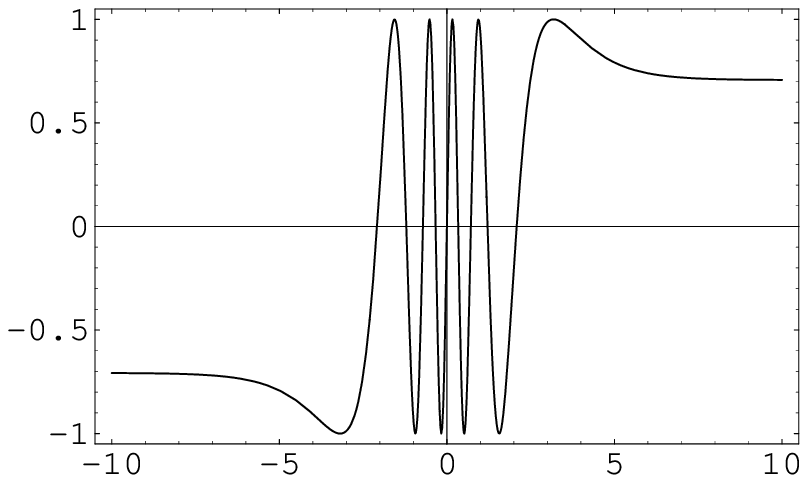}}
 \subfigure[$q=6$] {\label{fig_Fsinphib}
  \includegraphics[width=7cm,height=5cm]{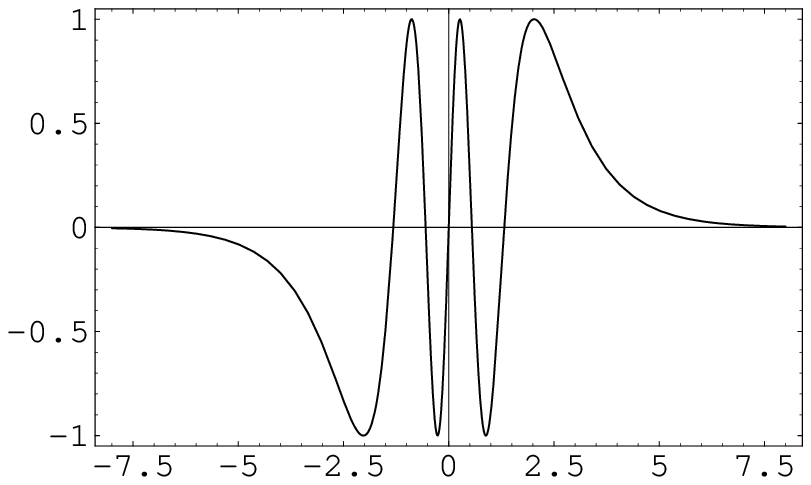}}
 \caption{The shape of $F(\phi)=\sin \phi$ for various values of
$q$ in $y$ coordinate. For $q=9.5$, which is between $4n$ and
$4n+2$, $F$ tends to a positive constant when $z\rightarrow
\infty$ and tends to a negative constant when $z\rightarrow
-\infty$. For $q=6$, which is an even number, $F$ tends to zero
when $z\rightarrow \pm\infty$.} \label{fig_Fsinphi}
\end{figure}

For the case $F(\phi)=\sin\phi$, the potential as a function of
$y$ for the left chiral fermions is
\begin{eqnarray}
  V_L(z(y))&=& -\frac{1}{2} \eta \cosh^{-1-2\tau} (ky)
      \bigg[\eta\left(\cos(2\phi(y))-1\right)\cosh ky    \nonumber  \\
     &&  +2kq\cos\phi(y)
         -\tau q \sin\phi(y)
          \sinh ky  \bigg].
\end{eqnarray}
For different values of $q$, $F(\phi(y))$ has different behaviors,
which should result in different types of the potential $V_L$.
According to the expression of $F(\phi)$:
\begin{eqnarray}
F(\phi(y))=\sin\left(2q \arctan \text{e}^{ky}  -\frac{\pi
q}{2}\right),
\end{eqnarray}
we have $F|_{y\rightarrow \infty} \rightarrow \sin
\frac{q\pi}{2}$. So, when $y\rightarrow \infty$, $F(\phi(y))$ has
different limits for different values of $q$:
\begin{eqnarray}
 F(y\rightarrow \infty) &>&0 ~~~~ \text{for} ~~ 4n<q<4n+2, \\
 F(y\rightarrow \infty) &=&0 ~~~~ \text{for} ~~~~~~~~ q=2n, \\
 F(y\rightarrow \infty) &<&0 ~~~~ \text{for} ~~ 4n+2<q<4n+4,
\end{eqnarray}
where $n$ is an arbitrary integer. The shapes of $F(\phi(y))=\sin
\phi(y)$ for various values of $q$ are shown in Fig.
\ref{fig_Fsinphi}.

Considering that $\tau$ does not change the characteristic of the
effective potentials acting on the left and right chiral fermions,
we mainly focus on the case $\tau=1$, for which one can get the
explicit forms of the potentials in $z$ coordinate
\begin{eqnarray}
 V_L(z)
  &=&\eta \bigg(
    \frac{k^2 z\sin\phi(z)}{(1+k^2 z^2)^{3/2}}
         +\frac{\eta \sin^2\phi(z)}{1+k^2 z^2} \nonumber \\
  &&-\frac{2kq\text{e}^{\text{arcsinh} kz} \cos \phi(z)}
         {(1+\text{e}^{2\text{arcsinh} kz})(1+k^2 z^2)}
  \bigg),   \label{VLz2}  \\
  V_R(z) &=& V_L(z)|_{\eta \rightarrow -\eta},  \label{VRz2}
\end{eqnarray}
where $\phi(z)$ is given by Eq. (\ref{phi_z}). The values at $y = 0$
are given by
\begin{equation}
V_L(0) =-V_R(0) = -k q \eta.
\end{equation}
Both potentials have the asymptotic characteristic:
$V_{L,R}|_{z\rightarrow\pm\infty}\rightarrow 0$. But for a given
coupling constant $\eta$, the values of the potentials at $z=0$
are opposite. It can be seen from (\ref{VLz2}) and (\ref{VRz2})
that the shapes of the potentials are determined by $\sin\phi(z)$
and $\cos\phi(z)$, which depend closely on the value of $q$.

\begin{figure}[h]
 \centering
 \subfigure[$0<q\leq 2, ~q=3$]{\label{figVLAa}
  \includegraphics[width=4cm,height=3cm]{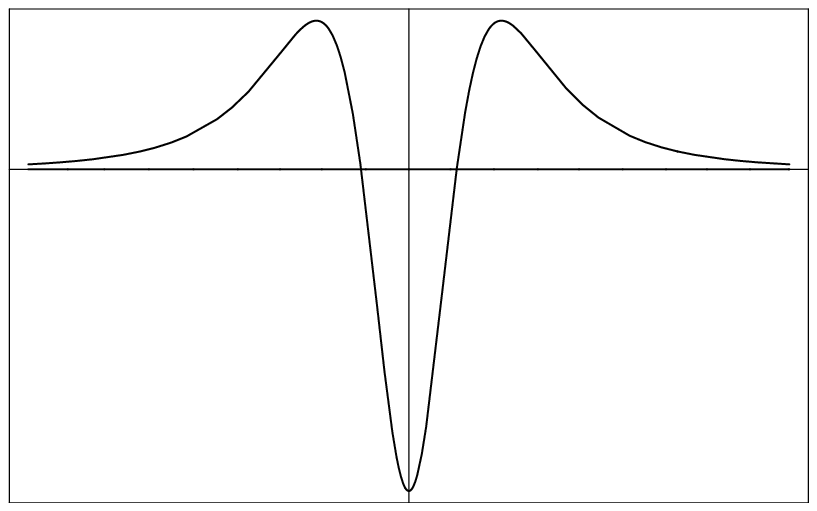}} 
 \subfigure[$4<q\leq 6, ~q=7$]{\label{figVLAb}
  \includegraphics[width=4cm,height=3cm]{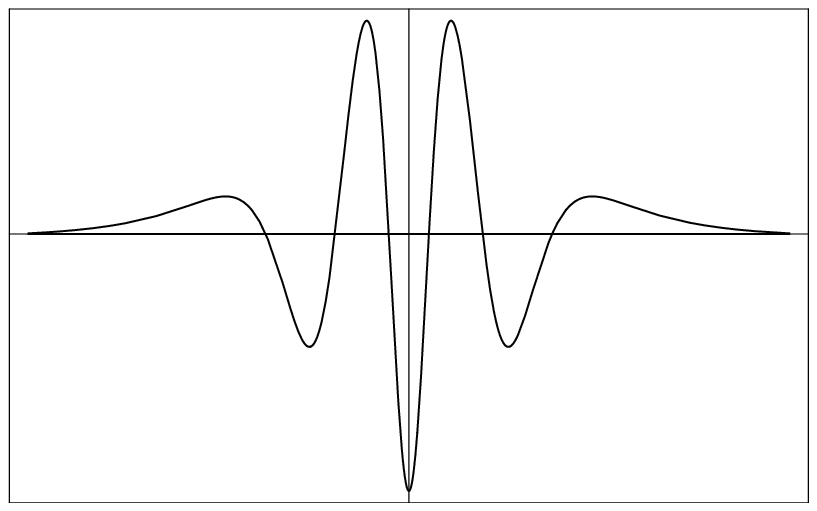}}
 \subfigure[$8<q\leq 10, ~q=11$]{\label{figVLAc}
  \includegraphics[width=4cm,height=3cm]{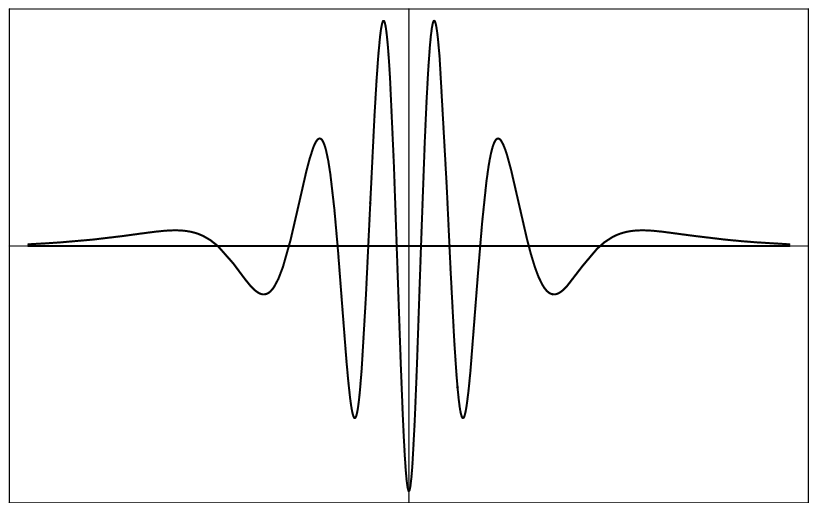}} 
 \caption{The shapes of the potential $V_L(z)$ \eqref{VLz2}  
  for $4n<q\leq 4n+2$ or $q=4n+3$.
  The potential has a negative value at the location of
  the thick brane, and $n+1$ finite positive barriers
  on each side which vanish asymptotically from above
  when far away from the brane.} \label{figVLzAbove}
\end{figure}

\begin{figure}[h]
 \centering
 \subfigure[$2<q\leq 4, ~q\neq 3$]{\label{figVLBa}
  \includegraphics[width=4cm,height=3cm]{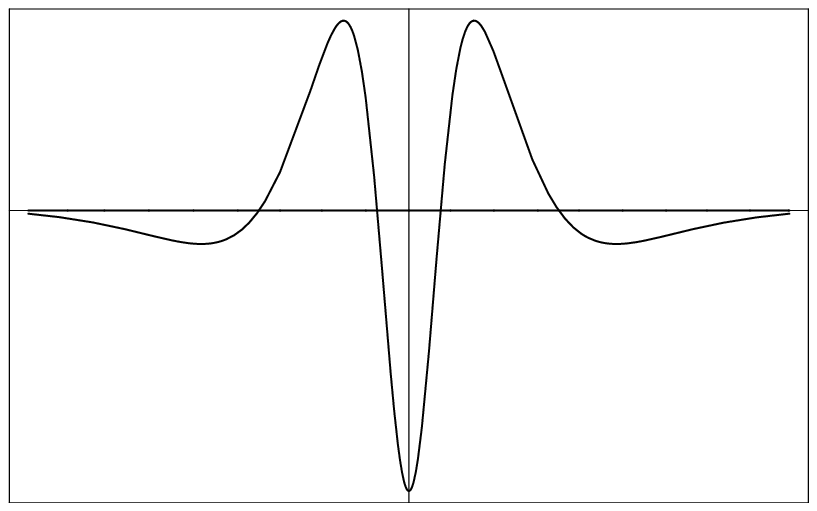}} 
 \subfigure[$6<q\leq 8, ~q\neq 7$]{\label{figVLBb}
  \includegraphics[width=4cm,height=3cm]{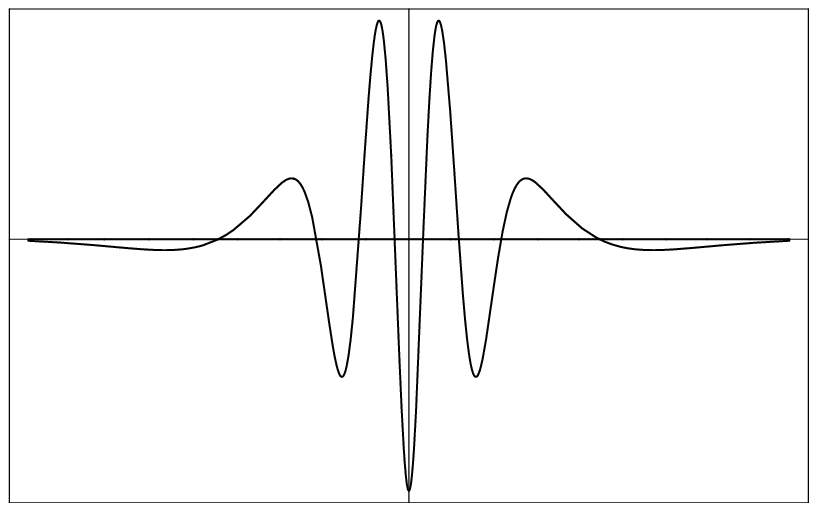}}
 \subfigure[$10<q\leq 12, ~q\neq 11$]{\label{figVLBc}
  \includegraphics[width=4cm,height=3cm]{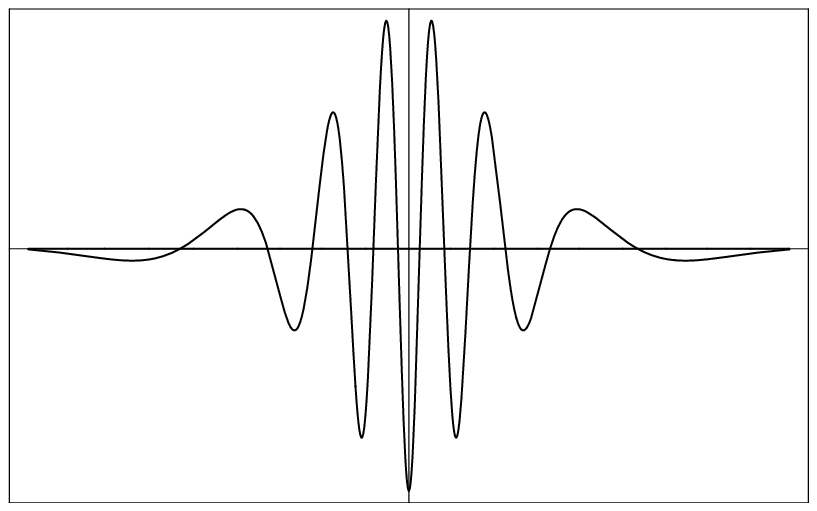}} 
 \caption{The shapes of the potential $V_L(z)$ \eqref{VLz2} 
  for $4n+2<q\leq 4n+4$ but $q\neq4n+3$.
  The potential has a negative value at the location of
  the thick brane, and $n+1$ finite positive barriers
  on each side which vanish asymptotically from below
  when far away from the brane.} \label{figVLzBelow}
\end{figure}

For general $q$, $V_L$ is not any more a volcano type potential.
Here we only discuss the case of positive $\eta$ and $q$, which
results in a negative potential at the location of the brane since
$k$ is positive. In order to localize fermions on the brane, we
also need at least a potential barrier on each side. In fact, for
$4n<q\leqslant 4n+2$ or $q=4n+3$, the potential for the left
chiral fermions has $n+1$ finite positive barriers on each side,
in which the last one on each side vanishes asymptotically from
above. The potential is always positive at long distances, so it
can trap the zero mode. The shapes of the potential for this case
are shown in Fig. \ref{figVLzAbove}. For $4n+2<q\leqslant 4n+4$
but $q\neq 4n+3$, the potential for the left chiral fermions has
also $n+1$ finite positive barriers on each side, but the last
barrier on each side vanishes asymptotically from below. The
potential is always negative at long distances, which indicates
that it can not trap the zero mode. See Fig. \ref{figVLzBelow} for
the shapes of the potential. Hence, in order to get a potential
for the left chiral fermions that can trap some fermion KK modes,
we first need the following condition
\begin{eqnarray}
 4n<q\leqslant 4n+2~(n\geq 0) ~~~\texttt{or}~~~ q=4n+3~(n\geq 0).
  \label{ConditionForVL}
\end{eqnarray}

\begin{figure}[h]
 \centering
 \subfigure[~$0<q\leq 2, ~q\neq 1$]{\label{figVRBa}
  \includegraphics[width=4cm,height=3cm]{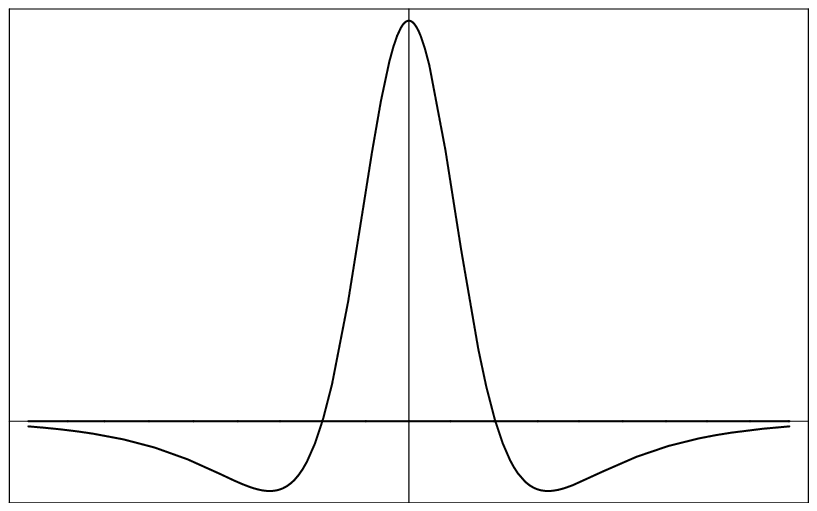}} 
 \subfigure[~$4<q\leq 6, ~q\neq 5$]{\label{figVRBb}
  \includegraphics[width=4cm,height=3cm]{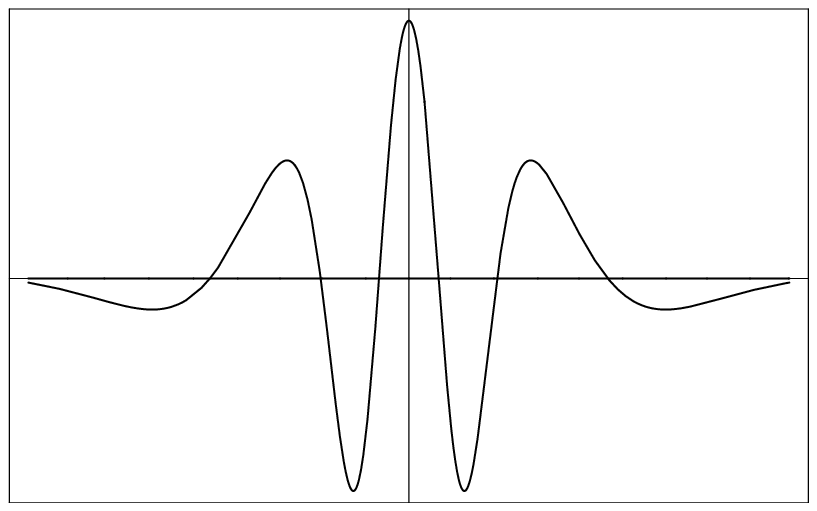}}
 \subfigure[~$8<q\leq 10, ~q\neq 9$]{\label{figVRBc}
  \includegraphics[width=4cm,height=3cm]{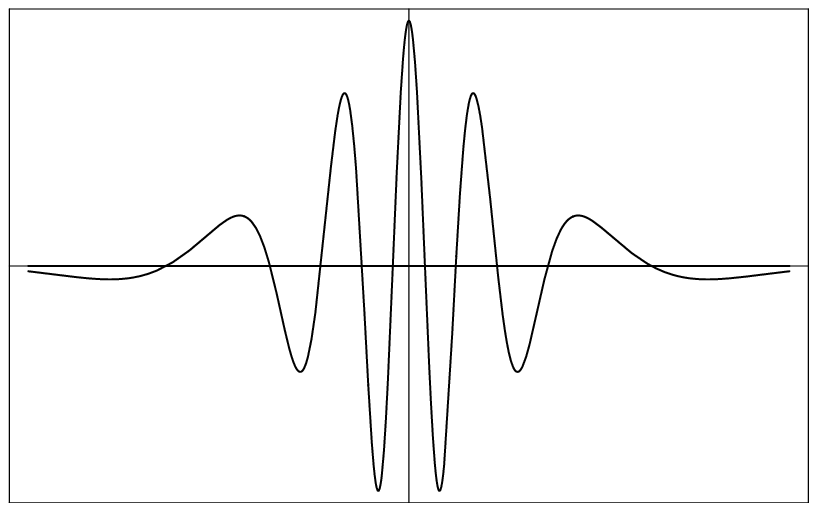}} 
 \caption{The shapes of the potential $V_R(z)$ \eqref{VRz2} 
 for $4n<q\leq 4n+2$ but $q\neq 4n+1$. The potential has a positive
value at the location of the thick brane, and $n$ finite positive
barriers on each side which vanish asymptotically from below when
far away from the brane.} \label{figVRzBelow}
\end{figure}

\begin{figure}[h]
 \centering
 \subfigure[$q=1$]{\label{figVRAa}
  \includegraphics[width=4cm,height=3cm]{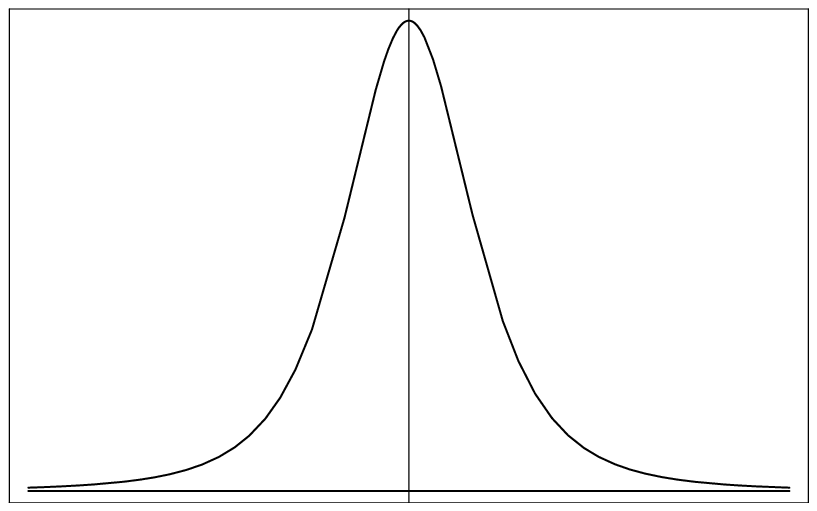}} 
 \subfigure[$2<q\leq 4,~q=5$]{\label{figVRAb}
  \includegraphics[width=4cm,height=3cm]{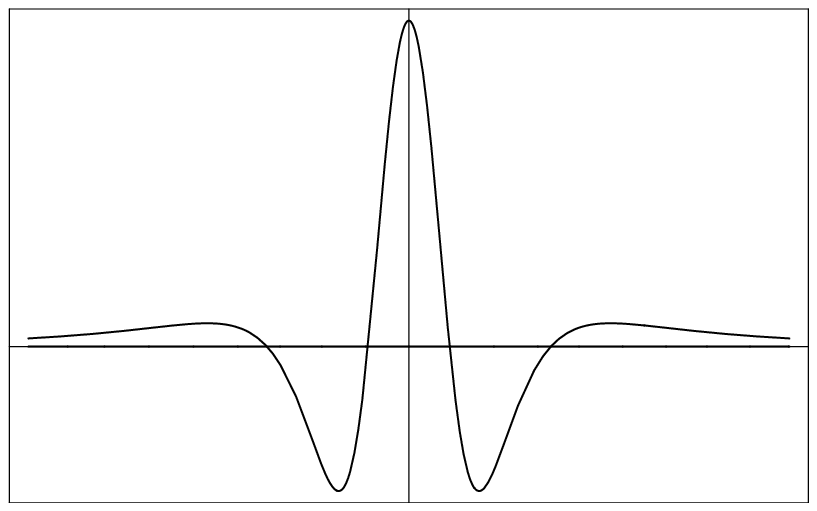}}
 \subfigure[$6<q\leq 8,~q=9$]{\label{figVRAc}
  \includegraphics[width=4cm,height=3cm]{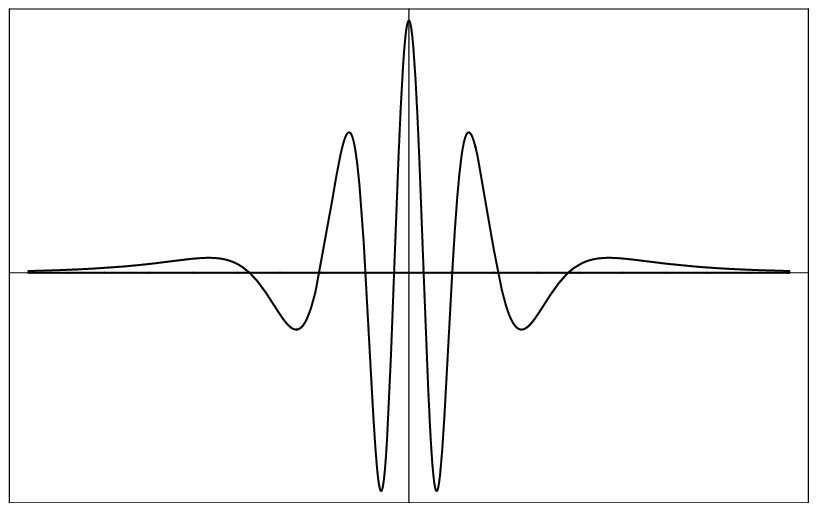}} 
 \caption{The shapes of the potential $V_R(z)$ \eqref{VRz2} 
 for $4n+2<q\leq 4n+4$ and $q=4n+1$. The potential has a positive value
at the location of the thick brane, as well as $n+1$ and $n$
finite positive barriers for $4n+2<q\leq 4n+4$ and $q=4n+1$
respectively on each side which vanish asymptotically from above
when far away from the brane.} \label{figVRzAbove}
\end{figure}

Next we discuss the relation of the potential $V_R(z)$ with the
parameter $q$. We also limit our discussion on positive $\eta$ and
$q$, which results in a positive potential at the location of the
brane for $V_R(z)$. This looks like that the potential could not
trap any KK modes of the right chiral fermions. But the result is
opposite. For $4n<q\leqslant 4n+2$ but $q\neq4n+1$, the potential
for the right chiral fermions has $2n+1$ finite barriers and
$2n+2$ finite wells, among them a positive barrier is located at
the location of the brane. The potential vanishes asymptotically
from below at long distances, which indicates that it can not trap
the zero mode. For $4n+2<q\leqslant 4n+4$, the potential has
$2n+3$ finite barriers and $2n+2$ finite wells. For $q= 4n+1$, the
potential has $2n+1$ finite barriers and $2n$ finite wells. For
both cases, the potential vanishes asymptotically from above at
long distances, which indicates that it can not trap the zero
mode. Hence, in order to get a potential for the right chiral
fermions that can trap some fermion KK modes, we need the
following condition
\begin{eqnarray}
 4n+2<q\leqslant 4n+4~(n\geq 0) ~\texttt{or}~ q=4n+1~(n>0).~
  \label{ConditionForVR}
\end{eqnarray}
This is a remarkable result which is very different from the case
considered in previous subsection, where the potential for the
right chiral fermions with positive $\eta$ and $q$ can not trap
any KK modes because it is always positive. The shapes of the
potential $V_R(z)$ for various values of $q$ are shown in Figs.
\ref{figVRzBelow} and \ref{figVRzAbove}.

Now we examine the zero modes for the left and right chiral
fermions. By setting $m_0=0$ and $F(\phi)=\sin\phi$, from Eq.
(\ref{CoupleEq1}) we find the left and right zero modes have the
following formalized solutions:
\begin{eqnarray}
 \widetilde{\alpha}_{L0}(z)
 &\propto& \exp\left(-\eta\int^z
  dz'\text{e}^{A(z')}\sin\phi(z')\right),
   \label{LeftZeroMode2} \\
  \widetilde{\alpha}_{R0}(z)
 &\propto& \exp\left(+\eta\int^z
  dz'\text{e}^{A(z')}\sin\phi(z')\right).
   \label{RightZeroMode2}
\end{eqnarray}
Using the same method in previous subsection, we can obtain the
restriction on the free parameters $\eta$ and $q$ from the
normalization condition (\ref{orthonormality}) for the zero modes
(\ref{LeftZeroMode2}) and (\ref{RightZeroMode2}). In $y$
coordinate, the restriction condition for the zero modes is
\begin{equation}
\int dy \exp\left(-A(y)-(\pm)2\eta\int^y dy'\sin\phi(y')\right)
  < \infty.
\end{equation}
When $y \rightarrow \infty$, we have $A(y)\rightarrow -ky$ and
$\sin\phi(y)\rightarrow \sin \frac{q\pi}{2}$, and so

\noindent $\left(-A(y)-(\pm)2\eta\int^y dy'\sin\phi(y')\right)
\rightarrow (k-(\pm)2\eta \sin \frac{q\pi}{2})y$. Thus, the
restriction condition reduces to
\begin{equation}
\pm 2\eta\sin(q\pi/2)>k   \label{condSinPhi}
\end{equation}
with ``$+$" for the left fermions and ``$-$" for the right ones.
For the left chiral fermions, combining (\ref{condSinPhi}) with
the constrain (\ref{ConditionForVL}) coming from the effective
potential $V_L$, the condition for localizing the zero mode is
turned out to be
\begin{eqnarray}
\begin{array}{c}
  4n<q<4n+2~~ \texttt{or} ~~q=4n+3~(n\geq 0), \\
  \eta>\frac{k}{2\sin(q\pi/2)}. \\
\end{array}  \label{condSinPhiLeft}
\end{eqnarray}
For the right chiral fermions, the localization condition of the
zero mode is
\begin{eqnarray}
\begin{array}{c}
  4n+2<q<4n+4~~ \texttt{or} ~~q=4n+1~(n\geq 0), \\
  \eta>\frac{k}{-2\sin(q\pi/2)}. \\
\end{array}\label{condSinPhiRight}
\end{eqnarray}
Note that the values $q=2n$ do not appear in
(\ref{condSinPhiLeft}) and (\ref{condSinPhiRight}). The reason is
that the value of $F(\sin\phi)$ will tend to zero at long
distances for $q=2n$, which results in the non-normalizable zero
modes. In Ref. \cite{MelfoPRD2006}, Melfo {\em et al } studied the
localization of fermions on various scalar thick branes. They
showed that only one massless chiral mode is localized in double
walls and branes interpolating between different $AdS_5$
space-times whenever the wall thickness is keep finite, while
chiral fermion modes cannot be localized in $dS_4$ walls embedded
in a $M_5$ space-time. In Ref. \cite{hep-th/0309162}, Bietenholz
{\em et al } investigated fermions in a brane world of the 3-d
Gross-Neveu Model and addressed in particular the question if
approximate chiral symmetry can come about in a natural way under
brane-type dimensional reduction. They found that a left-handed 2D
fermion localized on the domain wall and a right-handed fermion
localized on the anti-wall communicate with each other through the
3D bulk, and the two 2D fermions are bound together to form a
Dirac fermion of mass $m$. This involves a hierarchy problem with
respect to the fermion mass.

\begin{figure}[h]
 \centering
 \subfigure[Zero mode ($m^2=0,~q=9$)]{\label{figKKmodesSinPhia}
  \includegraphics[width=14cm,height=6cm]{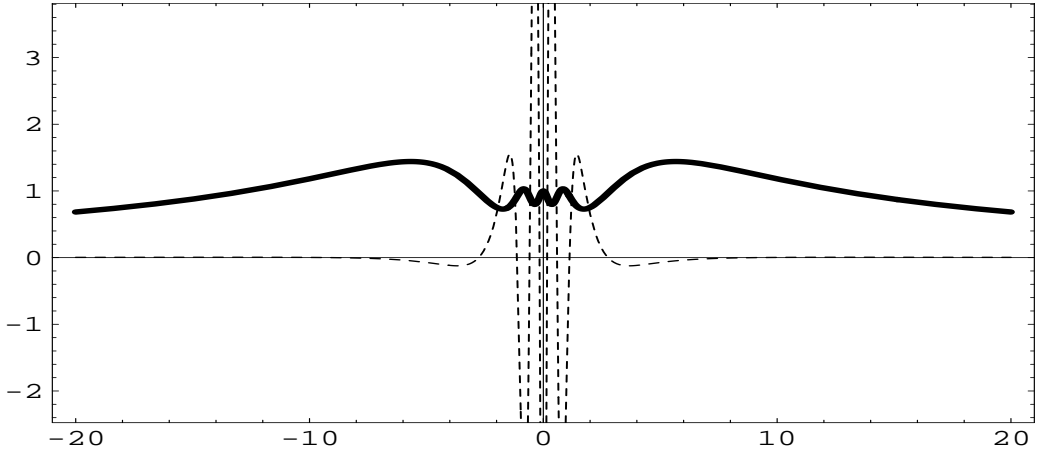}}
 \subfigure[Massive modes ($m^2=1$ for thick black
line, $m^2=10$ for thin gray line, $~q=9$)]
{\label{figKKmodesSinPhib}
  \includegraphics[width=14cm,height=6cm]{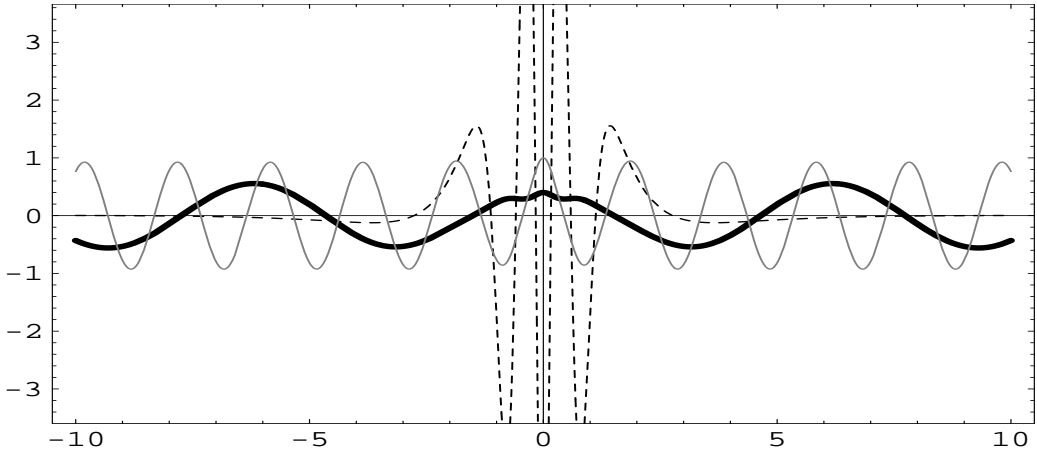}}
 \caption{The shape of the potentials $V_L(z)$ \eqref{VLz2}  
 (dashed lines),  the zero mode (\ref{LeftZeroMode2}) 
 and the massive modes for the left chiral
 fermions for the case $F(\phi)=\sin\phi$. The parameters are
 set to $k=1,q=9$, and $\eta=1$.} \label{fig_KKmodes_sinphi}
\end{figure}

For arbitrary $q\eta>0$, the two potentials suggest that there is
no mass gap but a continuous spectrum of KK modes. In Fig.
\ref{fig_KKmodes_sinphi}, we plot the massless and massive KK
modes for the left chiral fermions. It can be seen that the zero
mode is bound on the brane if the condition (\ref{condSinPhiLeft})
is satisfied. The massive modes with lower energy (especially the
zero mode) experience an attenuation due to the series of
potential barriers near the location of the brane.

To close this section, we make some comments on the issue of the
localization of fermions. Localizing the fermions on branes or
defects requires us to introduce other interactions besides
gravity. More recently, Volkas {\em et al } had extensively
analyzed localization mechanisms on a domain wall. In particular,
in Ref. \cite{Volkas0705.1584}, they proposed a well-defined model
for localizing the SM, or something close to it, on a domain wall
brane. There are some other backgrounds, for example, gauge field
\cite{LiuJHEP2007}, supergravity \cite{Mario,Parameswaran0608074}
and vortex background
\cite{LiuNPB2007,LiuVortexFermion,Rafael200803,StojkovicPRD},
could be considered. The topological vortex coupled to fermions
may result in chiral fermion zero modes \cite{JackiwRossiNPB1981}.

\section{Discussions}

In this paper, by presenting the shapes of the mass-independent
potentials in the corresponding Schr\"{o}dinger equations, we have
reinvestigated the possibility of localizing spin half fermions on
a thick brane for two kinds of kink-fermion couplings. It is shown
that, without scalar-fermion coupling, there is no bound state for
both the left and right chiral fermions. Hence, in order to
localize the massless and massive left or right chiral fermions on
the brane, some kind of Yukawa coupling should be introduced.

For the Yukawa coupling $\eta\bar{\Psi}\phi\Psi$, only one of the
potentials for the left and right chiral fermions has a finite
well at the location of the brane and a finite barrier on each
side which vanishes asymptotically. It is shown that there is only
one single bound state (zero mode) which is just the lowest energy
eigenfunction of the Schr\"{o}dinger equation for the
corresponding chiral fermions. When the condition $\eta q\pi>k$ is
satisfied, the zero mode is normalizable.

For the scalar-fermion coupling $\eta\bar{\Psi}\sin\phi\Psi$ with
$q>0$ and $\eta>0$, the potential for the left chiral fermions has
a finite well at the location of the brane as well as a series of
finite positive barriers on each side, and vanishes asymptotically
from above or below when far away from the brane. Under the
condition (\ref{condSinPhiLeft}), there exists a bound and
normalizable left chiral fermion zero mode.

It is worth to point out that, under the condition $q>0$ and
$\eta>0$, for the usual coupling $\eta\bar{\Psi}\phi\Psi$, the
potentials for the left and right chiral fermions have very
different shapes and only the left fermion zero mode could be
localized. However, for the coupling $\eta\bar{\Psi}\sin\phi\Psi$,
the potentials for the left and right chiral fermions have similar
shapes and the right fermion zero mode also could be localized on
the brane under the condition (\ref{condSinPhiRight}). The reason
is that, although the potential for the right chiral fermions has
a positive value at the location of the brane, it has some wells
and a series of positive barriers near the brane, which ensures
that it can trap the right chiral fermion zero mode on the brane.

Since the potentials for both scalar-fermion couplings vanish
asymptotically when far away from the brane, all values of $m^2>0$
are allowed, and there exists no mass gap but a continuous gapless
spectrum of KK states with $m^2>0$. The massive KK modes
asymptotically turn into continuous plane waves when far away from
the brane \cite{Lykken,dewolfe}, which represent delocalized
massive KK fermions.

\section*{Acknowledgement}

Y.X. Liu is really grateful to Prof. Jian-Xin Lu and Prof.
Huan-Xiong Yang for their invitation to visit USTC-ICTS and
interesting discussions. The authors are also thankful to Dr.
Wolfgang Bietenholz, Xin-Hui Zhang and Shao-Wen Wei for valuable
recommendations and discussions. L.J. Zhang acknowledges financial
support from Shanghai Education Commission. This work was
supported by the National Natural Science Foundation of China (No.
502-041016 and No. 10705013), the Doctor Education Fund of
Educational Department of China (No. 20070730055) and the
Fundamental Research Fund for Physics and Mathematics of Lanzhou
University (No. Lzu07002).

\end{document}